# Quantitative atom counting of Zn and O atomsby atomic resolution off-axis and in-line holography

U. Bhat[1], and R. Datta[1]

[1]*International Centre for Materials Science, Chemistry and Physics of Materials Unit, Jawaharlal Nehru Centre for Advanced Scientific Research, Bangalore 560064, India.*

**Abstract**

Quantitative atom counting of Zn and O atoms in zinc oxide(ZnO)epitaxial thin film by three different routes; reconstruction of phase from side and central band of atomic resolution off-axis and in-line electron holography are presented. It is found that the reconstructed phase from both side and central band and corresponding atom number for both Zn ($Z = 30$) and O ($Z = 8$) atom columns are in close agreementalong with the systematic increase in thickness for thinner sample area.However, complete disagreement is observed for thethicker sample area. On the other hand,the reconstructed phase obtained via in-line holography showsno systematic change with thickness.Phase detection limits and atomic model used to count the atoms are discussed.

Corresponding author e-mail address:

## I. Introduction

Phase is the fundamental information obtained by high resolution transmission electron interferometerexperiment[1].Phase shift encodes information on the potential of atomic ensembles and detailed knowledge of charge distribution which may be used to deducethe atomic arrangement and properties ofmaterials[2–4].Two established approaches,i.e., off-axis and in-line electron holography can be usedto retrieve phase information experimentally at atomic and sub-atomic length scale. In-line holography is popularly known as HRTEM (high resolution transmission electron microscopy). The first approach,which has the origin in Gabor's proposal of holographyand subsequent development of off-axis geometry by Leith and Upatnieks[5,6]. Off-axis geometry eliminates the twin image problem associated with the Gabor's original idea of in-line holography[6]. Gabor's proposalwas based on using a reference optical wavefront to interfere with the object wave,e.g., an electron micrographto overcome the resolution limit imposed by the geometrical aberrations of the electron lens. Such a hologram contains all the information about the object and the imaging system. Practical off-axis electron holography technique makes use of an electrostatic bi-prism for the electron interference developed by Möllensted and Düeker[7]. The two side bands (SBs) of the off-axis hologram containpure phase information. The central band (CB) is equivalent to inline holography with mixed amplitude and phase signals. Off-axis holography is a routine techniquefor medium resolution imaging of electric and magnetic fields[8–10]. Only recently, atomic resolution off-axis holography has been possible with the development of a special holography microscopeequipped withdouble bi-prism set up[11–14]. Double bi-prism set up eliminates Fresnel fringes and Vignetting effect essential for good quality atomic resolution off-axis hologram which usually has a small field of view[15,16]. Atomic resolution off-axis electron holography is a recent development where sub-atomic electron interference fringes encode phase information at that length scale.

On the other hand,reconstruction of phase from in-line holography requires series of imagesto be recorded at different focus values. Various reconstruction schemes for object exit wave(OEW) function have been developed from the experimental image series[1,17–20]. Development of both the experimental approaches to obtain phase information dates back to the BRITE EURAM program[21]. Comparisons of phase information by two different approaches have been performed by few groups both at medium and atomic scale resolutions.However, quantitative phase information obtained so far through off-axis and inline holography do not correspond to each other for the same sample area and depends on frequency range considered for the analysis[14,22–24]. Quantitative imaging is a recent area of active research inatomic resolution microscopy community[12,25–32]and understanding the accuracy on the experimental phase determination and its correlation with the property of materials is crucial for its success and contribution to material and microscopyscience as a whole. Both aberration corrected HRTEM and atomic resolution off-axis holography provide a unique opportunity to record phase information at the atomic and sub-atomic length scale.

In the present report, we compare the atomic scale phase information quantitatively by three different methods; off-axis electron holography using both SB and CB and in line holography. It is found that the peak phase values and corresponding atom numbers for both heavy Zn (Z =30) and light O (Z = 8) atoms are in close agreement between the SB and CB of off-axiselectron holography for thinner specimenarea with a systematic change in sample thickness. However, for thicker sample the agreement no longer holds. On the other hand, the phase information obtained via HRTEM method show a much lesser number of atoms than expected and does not change systematically with sample thickness.Phase detection limit in both the methods and atomic model used to count the atoms is discussed.

## II. Experimental details and data analysis

### A. Crystal growth

The ZnO epitaxial thin films were grown on '*c*' plane ZnO substrate under two different oxygen partial pressure ($p_{O_2}$) conditions using pulsed laser deposition (PLD) technique as described earlier[33,34]. Electron carrier concentrations can be controlled between $10^{19}$ to $10^{16}$ cm$^{-3}$ with $p_{O_2}$ $10^{-5}$ and $10^{-2}$ Torr, respectively. Though the original aim was to compare the difference in point defect distribution leading to change in carrier concentrations in these two samples, however, due to technical limitations at this point of time we restrict ourselves only to analyze atom counting by twodifferent phase contrast routes with sample thickness.

### B. Off-axis electron holography method, instrumentation and data analysis

The principle behind HRTEM and holography image acquisitionfor phase retrieval is shown schematically in Fig. 1 & 2.Atomic resolution off-axis electron holography is a recent development where electron interference fringesencode phase information at the sub-Å length scale where object wave is an atomic resolution electron micrograph. Double bi-prism set up atthe special location in the microscope column is important to avoid Fresnel artifact and Vignetting effect particularly at the atomic resolution where the field of view is severely restricted [Fig. 1 (c)][35]. The present data were acquired using aberration-corrected FEI TITAN 80-300 Berlin holography special TEM operated at 300kV in adouble bi-prism setup.Through focalimage series was acquired at a focus range of -10 to +10 nm with $\Delta f$ 1 nm. Third order spherical aberration coefficient ($C_S$)was set close to zero. It was already mentioned before that the aberration correction improved the phase detection limit by a factor of 4, i.e., $2\pi/20$ to $2\pi/80$[11]. Through focal holography method provides extraction of phase through CB using standard algorithm used for HRTEM, in the present case combination ofPAM (Paraboloid method) and MAL(Maximum-likelihood). MAL corrects exit wave function iteratively,based on a least square formalism. Series of images improves the signal to noise ratio significantly in the phase detection from the SB reconstruction using the Berlin code[14].Earlier comparison of phase values based onthe medium resolution reported poor signal to noise ratio for a single image SB reconstruction[24]. The details of the principle behind the method can be found inref. [14]and is shown schematically in Fig. 2 (a). Example FFT of the atomic resolution hologram from ZnO is shown in Fig. 2 (b). The cut off frequency for CB and SB are 14 and 12 $nm^{-1}$, respectively. The cut off frequencies are chosen in such a way that it does not overlap with the neighboring band.

### C. In-line holography method and data analysis

HRTEM data was acquired in a double aberration-corrected FEI TITAN 80-300 kV transmission electron microscope available at ICMS, JNCASR, Bangalore. An optimized phase contrast transfer function (PCTF) with $C_s$= -35 µm, $f$ = 8 nm and a point to point resolution better than 0.8 Å at 300 kV was set for the experimentation. Image series were recorded under these conditions; $C_s = -35\ \mu m$ and focus range -10 to 10 nm with $\Delta f$ 1 nm with the exposure time 1s. However, only 10 numbers of images are contemplated for the reconstruction. We did not observe any difference in the reconstructed phase image between 10 & 20 number of imagesconsidered for the reconstruction. The image series was reconstructed using the Gerchberg-Saxton scheme as implemented in MacTempas. Following are the parameters employed for the reconstruction; $C_s = -35\ \mu m$, acceleration voltage 300kV, area of reconstruction 1024×1024 (pixel), and objective aperture size $g_{max}$=2 Å$^{-1}$, neither any phase plate nor any filter is used. Strong central beam condition is considered. The phase image obtained was further corrected for the residual aberration using the digital aberration correction scheme available within the package. Example images before and after aberration correction are shown in Fig. S1 and the schematic of reconstruction steps are shown in Fig. 2 (e).

### III. Results and discussion

### A. Phase detection limit

Resolution is the most important parameter in high resolution transmission electron microscopy. In the presence of aberration, the point resolution $g_S$ is defined by the first zero crossing of the phase contrast transfer function (PCTF) on the frequency axis under optimum $C_s$ and defocus $\Delta f$ [Fig. S2]. The information limit $g_i$ of a microscope is the maximum information which can be transferred and isdefinedby the last point of the PCTF functionjust above the noise level and usually damped by various incoherent aberrations. The information encoded between the point resolution and information limit is not directly interpretable. For example, in an aberration corrected microscope one can obtain resolution $g_S$ better than 0.8Å, which is sufficient to resolve any chemical bonds in the crystalline material along high symmetry orientation. This reveals the structure of the material in terms of periodic arrangement of atoms.

Similar to resolution, minimum detectable amplitude and the phase signal of an electron waveafter interacting with the specimenpotential is equally important to evaluatethe smallest gradient of electric and magnetic fields, distinguishing atoms between the columns and counting atoms along the columns. Below is the brief discussion on phase detection limit in both off-axis and in-line holography in the context of present data.

In electron holography, following the procedure described by Lichte[36], the phase detection limit in a medium resolution hologram is given by

$$\sigma_\varphi = \frac{\sqrt{2e}}{p\sqrt{V^2 j_0 \tau STE(u_c)}} \tag{1}$$

Where, $e$ is the charge of the electron, $V$ is the fringe visibility, $STE(u_c)$ is the signal transfer efficiency of the CCD camera and $j_0$ is the current density during the exposure time $\tau$ over the area $p^2$.

Lichte has shown that the phase detection limit improves with increasing electron dose $N$nm$^{-2}$andincreasing lateral resolution $p$ of reconstructed wave.For atomic resolution holography as the width of hologram ($w_{hol}$) is related to the resolution $q_{max}$ ($w_{hol} \geq 4\,psf$), where $psf$ is the point spread function of the electron microscope.The above equation can be modified for a C$_s$ corrected microscope to

$$\delta\varphi_{lim} = \frac{4\sqrt{\pi}\ snr\ C_s}{|\mu|.V_{inel}.V_{inst}.V_{MTF}\sqrt{-\ln(|\mu|)\frac{B_{ax}}{ek^2}\epsilon t DQE(q_c)}} \times \frac{q_{max}^4}{k^3} \quad (2)$$

Where, *snr* is the signal to noise ratio, $q_{max}$ is the resolution in reciprocal space, $\mu$ is the degree of spatial coherence, $V_{inel}$, $V_{inst}$, $V_{MTF}$ are the hologram contrast arising due to inelastic scattering, instabilities and Modulation Transfer Function (MTF) of the CCD respectively. $DQE(q_c)$ is the Detection Quantum Efficiency of CCD camera, $B_{ax}$ is the brightness of the electron source, electron wave number *k* and e is the charge of the electron[11].

In the present experimental hologram with fringe spacing (*s*) of 0.0469 nm, the phase detection limit is 0.00023rad for an area $p^2 \sim 100$ nm$^2$ (512×512 pixels), V = 15%, and electron dose 16×10$^6$ nm$^{-2}$, which is the area of reconstruction in the present case. At the limit of resolution where the lateral resolution of wave should be selected as 4 times *psf*, i.e., for *p* =0.32 nm, phase detection limit is 0.007365 rad. With this phase detection limit counting of incremental atoms both for O (0.109 rad) and Zn (0.284rad) atoms is possible. The dependence of theoretical phase detection limit on V, electron dose and lateral resolution are given in supplementary[Fig. S3]. It can be seen that the phase detection limit changes within the same order of the magnitude with some variation in V, p and N thus should not affect the atom counting both in the case of Zn and O. In this context, Lehman et al [11] reported phase detection limit 2π/80 for an aberration corrected holography microscope. Cooper and Voelkl improved the phase detection limit to 0.001 and 2π/1000 (0.00628) by long exposure and multiplicity of holograms along with bi-prism and sample drift correction, respectively[37,38]. However, none of the latter two cases above used double bi-prism set up which eliminates Fresnel fringe and improves the phase detection limit significantly.

On the other hand, in the context of HRTEM, phase detection limit has not been discussed. Experimentally, distinguishing between B and N atoms has been reported with peak phase values as 0.022 and 0.026 rad, respectivelywith a difference of 0.004rad between the two atoms[39] [Fig. S4]. It is the shape and contrasts both responsible for the detection of atoms. The peak phase value on the atom position depends on atomicscattering and structure factors, microscope transfer function, and resolution. This will be reflected in the recorded image intensity as well. The changes in peak values for both phase and intensity can be calculated theoretically [see section II.B.]. However, there is another factor, i.e., the standard deviation in the vacuum phase value from reconstruction methoddetermines the experimental phase detection limit. Experimentally, it is the standard deviation of intensity and reconstructed phase in the vacuum will limit the interpretable phase change, i.e., typically 0.023 rad from the present result.In case of in holography,the number is better, i.e., 0.00488 rad (see also section III).

### B. Atomic potential model

It is necessary to compare the results with the theoretical reference values to quantify the atom numbers from the reconstructed phase shift. This method involves modeling the atomic potentialas imaging electron directly interacts with it giving rise to what is called object exit wave (OEW) function.Moreover, the lens phase contrast transfer function (PCTF) and aperturediameter ($k$ in Å$^{-1}$) modify the phase of the OEW further on the way to the recording device. The size of the nucleus (1.6 to 15 *fm*) is extremely tiny compared to the size of the atoms consisting of nucleus and surrounding electron clouds (0.1 to 0.5 nm). For a stationary atom, the Coulomb potential is $\propto \frac{1}{r}$, and there is a singularity at the center of the atom. The imaging electrons mostly see the nuclear potential, and the surrounding electrons shield the effect[40]. Inelastic events are negligible compared to elastic events(imaging electrons) for thin sample. Various theoretical atomic potential models are available in the literature[40–42]. In the present investigation, Hartree-Fock atomic model projected along the $z$-direction is considered which is given by

$$v_z(x,y) = \int_{-\infty}^{+\infty} V_a(x,y,z)dz$$

$$= 4\pi^2 a_0 e \sum_{i=1}^{3} a_i K_0(2\pi r\sqrt{b_i}) + 2\pi^2 a_0 e \sum_{i=1}^{3} \frac{c_i}{d_i} \exp(-\pi^2 r^2/d_i) \tag{3}$$

with $r^2 = x^2 + y^2$

Where, $a_0$ is the Bohr radius, $a_i, b_i, c_i, d_i$ are the parameterized coefficients. $K_0(x)$ is the modified Bessel function of order zero[19].

The projected atomic potential of Zn and O atoms calculated by the above equation is given in supplementary [Fig. S5]. The potential function is asymptotic due to $\frac{1}{r}$ dependence. Therefore, it is necessary to consider inner and outer bound of the potential while calculating the phase shift and image of the atom. The atomic scattering factor $f_e(k)$ (according to Moliere) and image of the atoms depends on the inner and outer cut off potential [Fig. S6]. However, it is observed that there is a limit in both inner and outer cut off, beyond which the change in $f_e(k)$ or peak values of image of the atom do not change significantly. For the present report, the limits selected are 0.01 and 1 Å, for calculating images of isolated Zn and O atoms using standard formula [See supplementary for more information].

The image of single isolated atom based on the model potential above, equation (3)can be calculated directly using electron scattering amplitude as given by the following equation;

$$g(x) = \left|1 + 2\pi i \int_0^{k_{max}} f_e(k) \exp[-i\chi(k)] J_0(2\pi kr) k dk\right|^2 \tag{4}$$

Where, $f_e(k)$ is the electron scattering factor in the Moliere approximation using the projected atomic potential. $\chi(k)$ is the aberration function, $k_{max} = \alpha_{max}/\lambda$ is the maximum spatial frequency in the objective aperture and $J_0(x)$ is the Bessel function of order zero.

The effect of inner and outer bound of potential and the resolution of the microscope as set by the objective aperture diameter on the shape and peak values of single atomphase shift and image intensity are given in supplementary[Fig. S7 and S8]. The real part of the wave transfer function,$\cos(\chi)$is neglected (i.e. set to zero) and imaginary part$\sin(\chi)$, which is the pure phase partis set to 1 tomimic Scherzer like transfer function for the atoms in a periodic lattice within weak phase object approximation. Similar to the scattering factor, the peak value of phase and intensity do not change significantly below an inner cut off of 0.01Å, and no significant change is observed with the outer bound. This is true in case of both Zn and O atoms. The peak phase and intensity values changes with the size of the aperture [Fig. S9]. In the present case an aperture size of 2Å$^{-1}$is used.

Two different theoretical phase values; peak and mean for a given atom which can be considered to interpret the reconstructed phase for counting the number of atoms. However, atoms are never stationary in the crystal and due to finite temperature atoms oscillate (0.0073and 0.0072 Å, for Zn and O atoms in ZnO at 293K [43]) about the mean position. Therefore, an incoming probe electron sees a blurred atom position. Aberration of the microscope will cause further blurring. However, the amplitude of thermal vibrationat room temperature and resulting blurring is smaller compared to the blurring due to aberration and is not considered in the present investigation. By numerical evaluation, one can find that the peak phase shift values have a coarse dependence of $Z^{0.6-0.7}$ and deviation can be observed due to valence electron filling with the atomic number[44]. On the other hand, mean phase shift value is sensitive to the inner and outer bound of potential [Table. S1]. Mean phase shift value does not change strongly with the inner cut off for less than 0.01 Å but changes significantly with outer cut off for less than 1 Å of the potential. But beyond 1 Å, it does not change significantly. An inner cut of 1pm and outer cut off of 50/25 pmcorresponding to experimental size of the Zn and O atoms, respectively considered for extractingmean phase shift value.The theoretical mean phase shift value is calculated bythree-dimensional integration of the potential and dividing with the volume bounded by the limits. The peak phase shift values for a microscope resolution of 0.8 Å obtained from literature and multislice calculation as implemented in MacTempas for the atoms in a crystal along with mean phase shift values are given in Fig.3.The two curves corresponding to peak phase values match well for fewer atoms in a column but deviates from linearity due to dynamical scattering for a higher number of atoms. The mean phase is found to be smaller (~ factor of 0.5) compared to the peak phase value, and this has implications on the atom number assignment by two different reference parameters and is discussed next.

## C. Atomic resolution off-axis electron holography

In this section the experimental phase information retrieved from both SB and CB off-axis hologram ofZnO film with varyingthickness is analyzed. ZnO films with two different thickness along <11-20> and <01-10> orientations are considered.The extinction lengths ($\xi_g$) are108.6 and 142.4 nm for <11-20> and <01-10> Z.A., respectively. Fig.4&6are the amplitude and phase images corresponding to CB and SB obtained for the two different areas marked as P and Q. The peak phase values on top of Zn and O columns have been evaluated, and selected columns at three different distances corresponding to different thickness levels from the edge of the specimen for area P are indicatedin the figures. Same columns are considered for the comparisons of two different OEW reconstructed using theCB and SB. The atom numbers corresponding to Zn atom evaluated from the peak and mean phase values are plotted in Fig. 5 for area P.Difference in the number of Zn atoms between CB and SB is within ±1 and ±3 corresponding to reference peak and mean phase respectively.Another noticeable point is that similar amount of Zn and O atoms are obtained for different areas for area P, suggesting that peak phase values as used from the theoretical model fits well in this case for both light and heavy atoms adjacent to each other. The reconstructed phase values and corresponding atom numbers for both Zn and O atoms in the neighboring columns for area P are given in Fig. 6. It can be seen that the atom numbersare in close agreement with difference ±1 atoms for Zn and O atoms.A Similar comparison is given forarea Q in Fig 7 shows the reconstructed phase and amplitude images. Fig 8 shows a comparison of phase shift as well as the atom numbers for Zn between the CB and SB. It can be seen that there is a gradual increase in the atom number with the thickness for the SB but almost constant (but different than the SB) atom number is obtained from the CB. Thus, for area Q the match is poor between CB and SBbecause of relatively thicker sample area and stronger dynamical effect.

### D. Inline holography/HRTEM

Fig.9 shows the reconstructed phase image of ZnO film along <11-20> orientation from different thickness regions of the sample. The peak phase values from Zn columns are given in the line scan for the columns indicated in the image. The peak phase values remain almost same between thinner and thicker regions of the sample at around 0.18rad which corresponds to~1Zn atom. In case of O peak phase value is around 0.09 rad and corresponds to ~1 atom.

Lehmann *et. al.*[14] first described the difference between in-line and off-axis electron holography at atomic resolution in GaAs crystal along <1-10> Z.A. The phase and amplitude reconstructed from the SB and CB agrees well up to a thickness of 3/2 times the extinction length, but significant deviations observed at lower frequency and thicker specimen area. The agreement between the two methods for the thinner area, is due to the similar wave function reconstructed in the limit of linear imaging with negligible inelastic scattering. However, for thicker area, due to significant inelastic scattering, reconstruction methods corresponding to CB and SB yield two different wave functions. This is because the mathematical formulation of SB contains average OEW function, while CB contains sum of squared OEW function. It is mentioned that the deviation observed in the thicker area between CB and SB reconstructed wave function maybe either due to fundamental quantum mechanical differences or numerically difficult inversion of the imaging process. In the present investigation, our results agree with Lehmann *et. al.* observation for the CB and SB reconstruction of off-axis experiment. However, observation made in HRTEM experiment is not comparable with the CB reconstruction. This could be because of the reconstruction scheme employed in the MacTempas package.

Counting of atoms depends on the theoretical reference phase values, i.e., mean or peak values of phase. We obtain ~ 3 times higher atom numbersfor Zn and O using reference mean phase value compared to peak phase value. The experimentally observed higher mean phase values compared to theory could be because of incoherent aberrations or vibrations present in the recorded image.

## IV.     Conclusions

In conclusion, atomic resolution reconstructed phase of Zn and O atoms in ZnO epitaxial thin film is compared between off-axis and in-line holography techniques. While holography method has an excellent match in atom numbers for both Zn and O atoms extracted from SB and CB for thin sample area,however, for thicker sample the atom numbers do not match. In case of in-line holographic reconstruction of HRTEM data, the atom number do not change systematically with increasing sample thickness, and a constant atom number of one is obtained throughout the reconstructed area.

**Acknowledgement**

R. Datta sincerely thanks Prof. Micheal Lehmann and Dr. Tore Niermann for hosting and assisting in the acquisition of atomic resolution off-axis holograms and data analysis. R. Datta also thanks DFG and ICMS for the funding.U. Bhat sincerely acknowledges JNCASR and ICMS for the financial support.

**Supplementary material**

Supplementary material contains information on example HRTEM images with and without digital aberration-correction, PCTF, phase detection limit in off-axis holography, atomic potential model and corresponding images and exit wave function.

**Figures:**

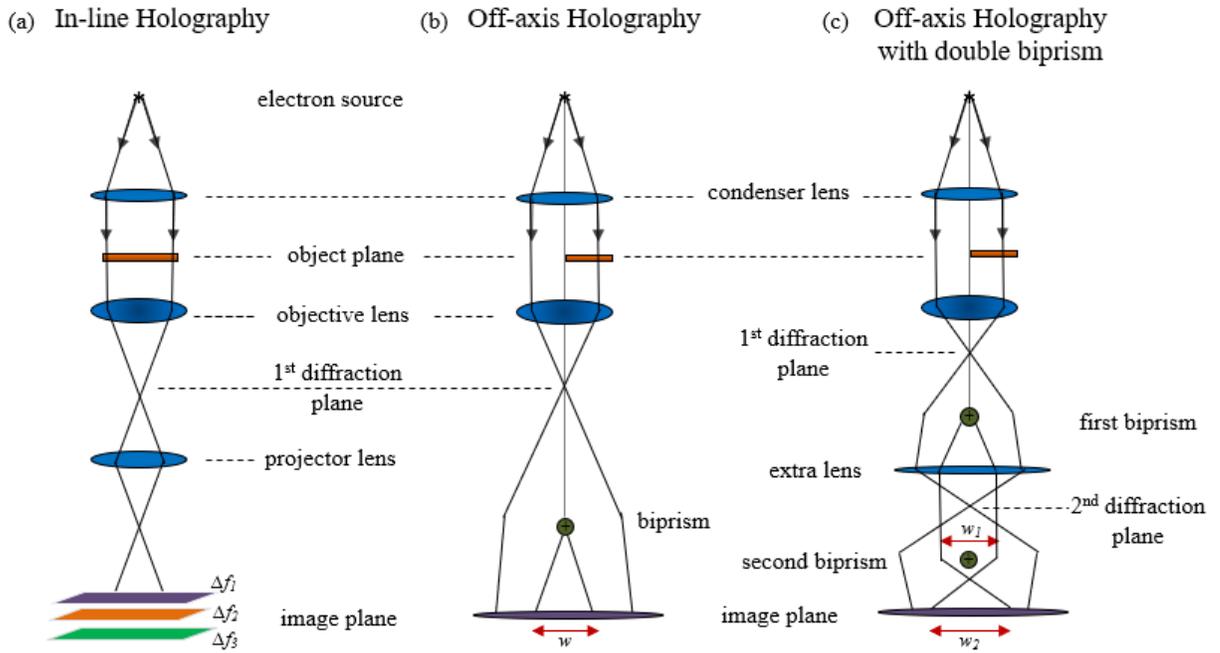

**Figure 1.** The schematics showing principle of image formation for (a) HRTEM, (b) off-axis electron holography, and (c) off-axis holography with double bi-prism set up.

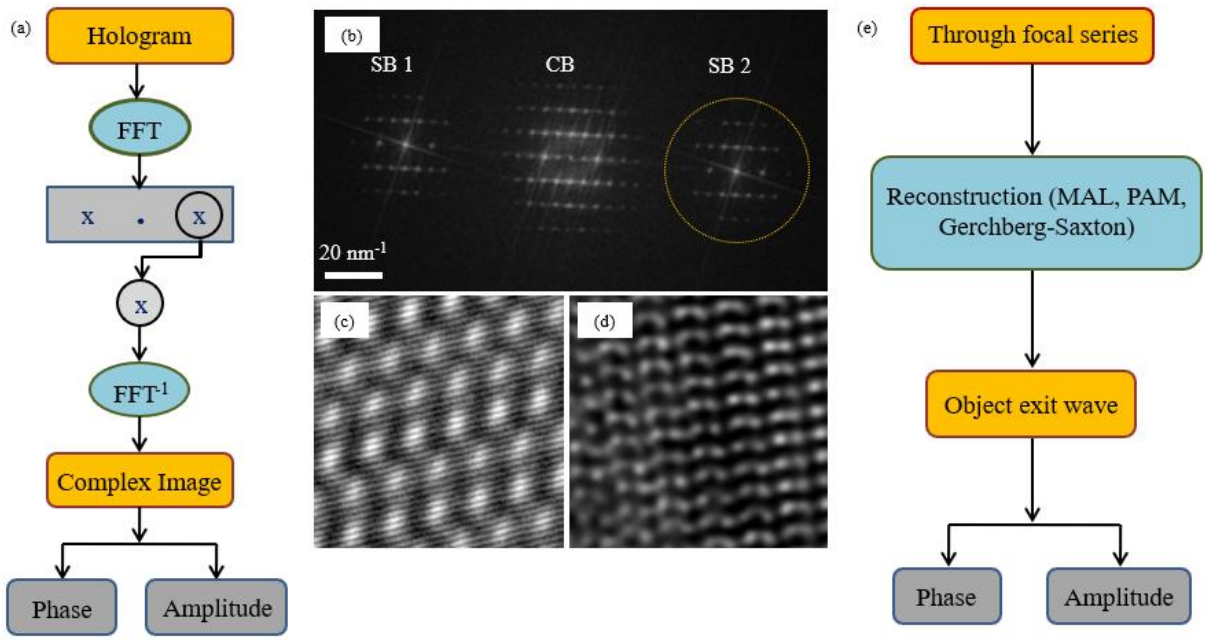

**Figure 2.** (a) & (e) Steps involving in reconstruction methods to extract phase and amplitude from the hologram and HRTEM image series, respectively. (b) Fourier transform of the hologram showing one CB and two SBs. (c)& (d) are the example atomic resolution hologram and HRTEM image of ZnO epitaxial thin film along <11-20> Z.A..

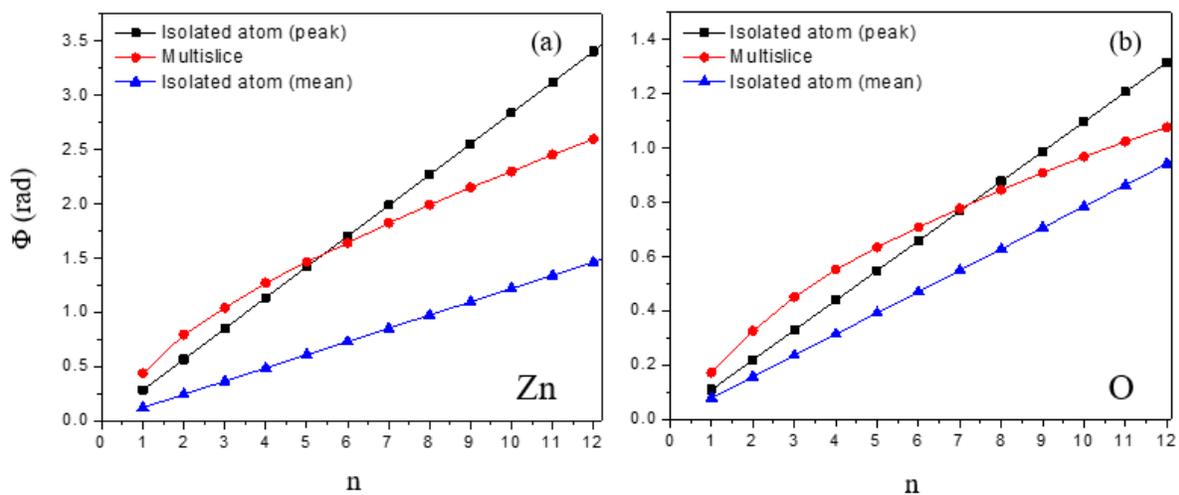

**Figure 3.** Peak phase shift of (a) Zn and (b) O atoms with increase in the number of atoms in the column calculated using isolated atom model and multislice method considering dynamical scattering. The resolution was set to 0.5 Å. Also plotted mean phase shift with the inner and outer bound of potential 10 to 50 pm, respectively.

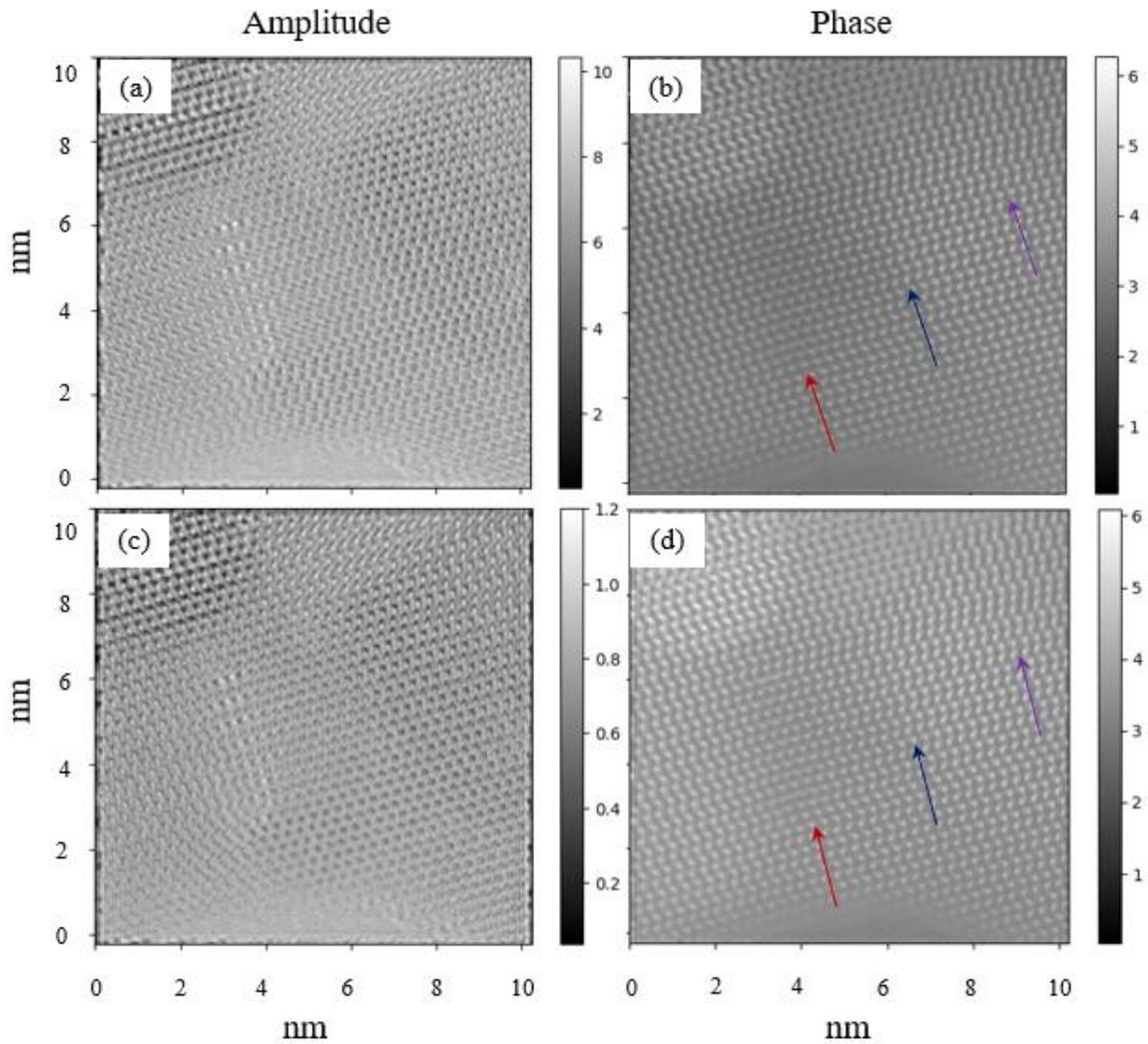

**Figure 4.** (a) & (c) Amplitude and (b) &(d) phase image of ZnO along <11-20> Z. A for area P obtained through reconstruction CB and SB of off-axis electron hologram, respectively. The corresponding complex wave function can be found in the supplementary Fig.S10[(a) and (b)].Three different arrows are indicated in the phase image along which the peak and mean phase values are extracted. Larger dots and smaller dots are corresponding to Zn and O atoms, respectively.

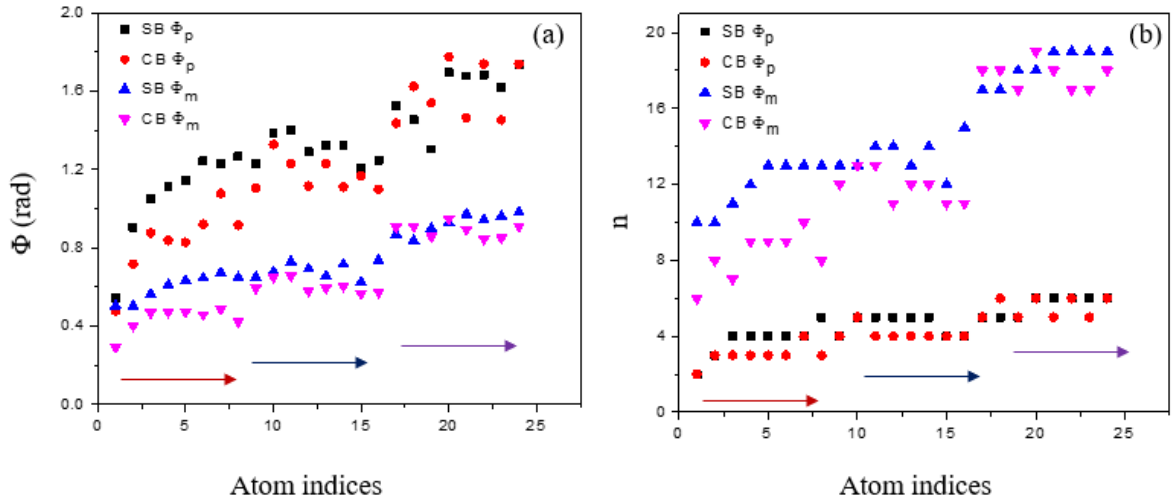

**Figure 5.** (a) Peak ($\Phi_p$) and mean ($\Phi_m$) phase shift and (b) corresponding atom numbers of Zn atom along three different arrows from area P reconstructed from SB and CB. Zn atom number matches well between SB and CB with ±1 atom. However, the number of atoms derived from the mean phase value is three times higher than peak phase value.

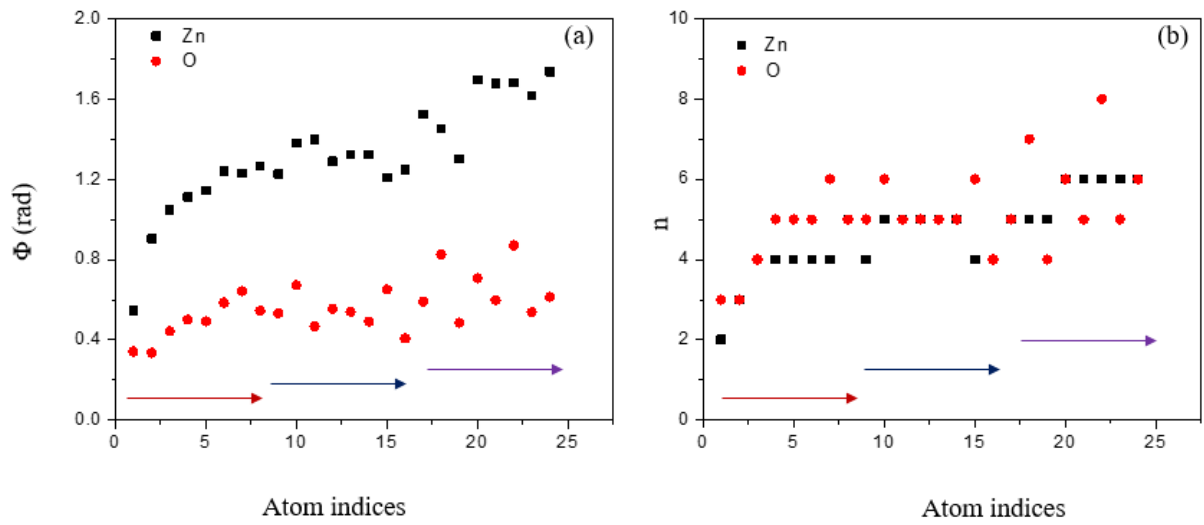

**Figure 6.** Comparison of (a) peak phase shift and (b) corresponding atom number with variation of thickness in Zn and O columns for area P. Almost similar number of atoms are obtained for Zn and O atoms at the neighboring sites.

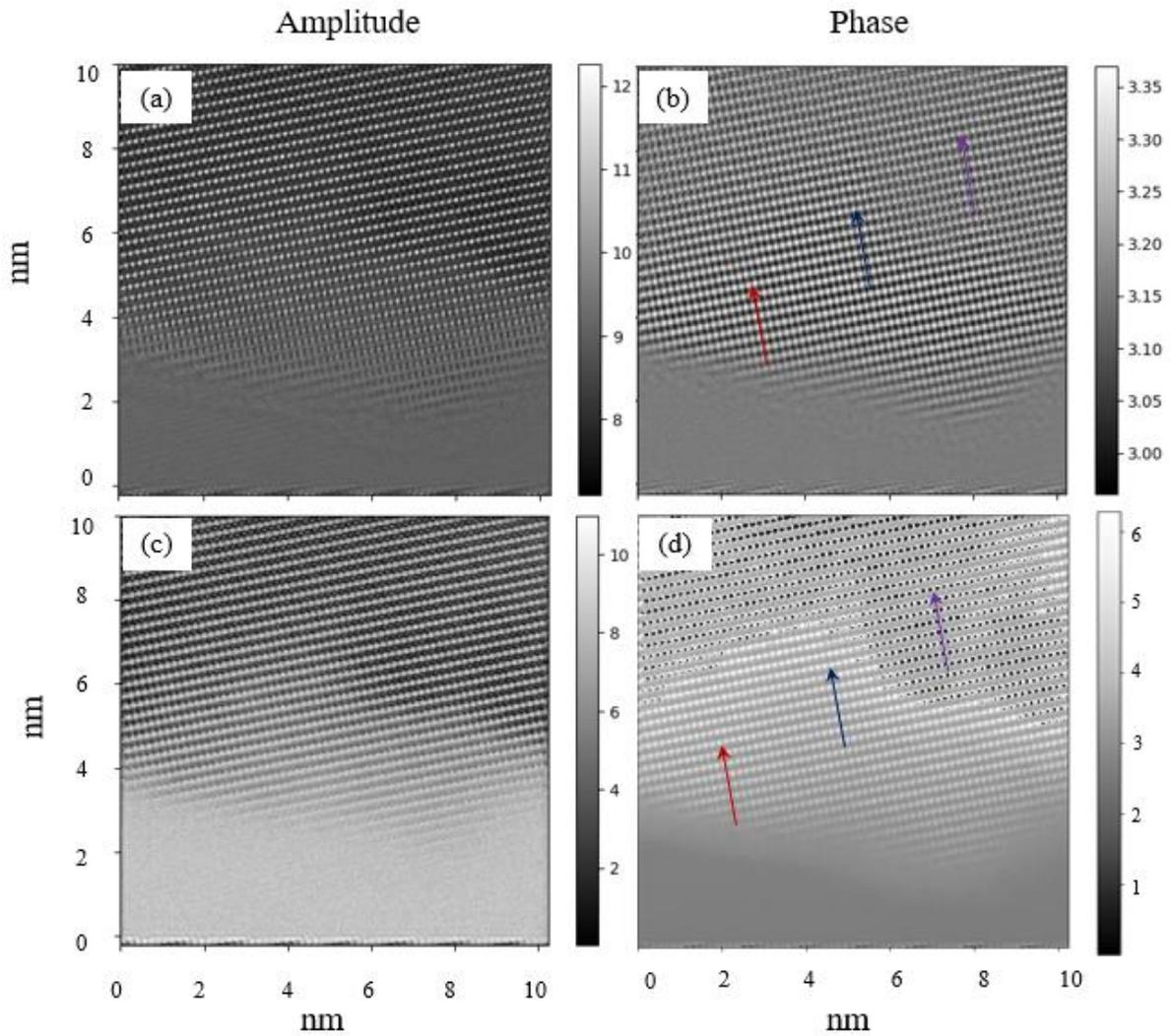

**Figure 7.**(a) & (c) Amplitude and (b) & (d) phase image of ZnO along <01-10> Z. A. for area Q obtained through reconstruction CB and SB of off-axis electron hologram, respectively. The corresponding complex wave function can be found in the supplementary Fig.S10[(c)and (d)].Three different arrows are indicated in the phase image along which the peak and mean phase values are extracted.

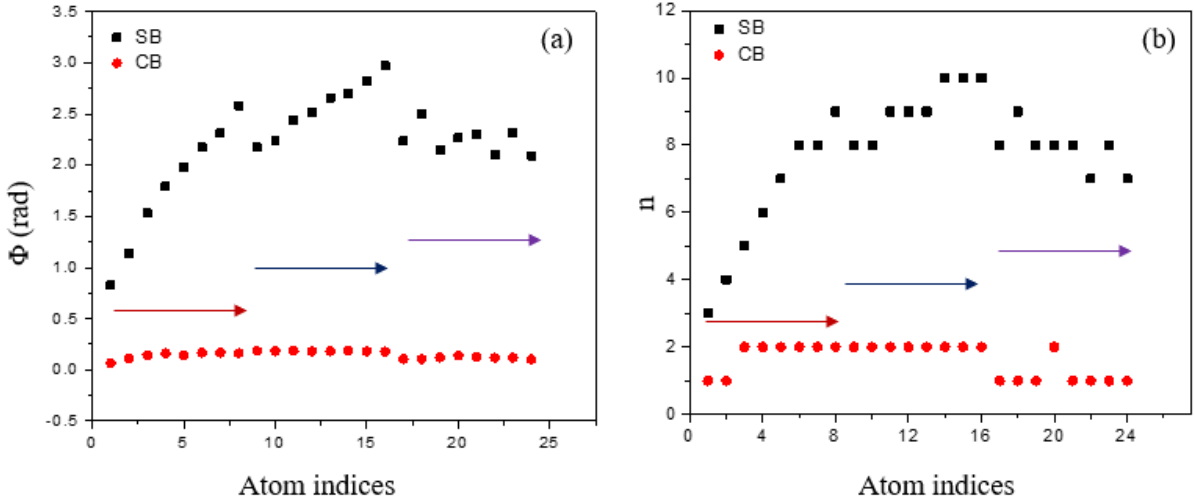

**Figure 8.** (a) Peak ($\Phi_p$) phase shift and (b) corresponding atom numbers of Zn atom along three different arrows from area Q reconstructed from SB and CB. No agreement is found in the number of Zn atoms between SB and CB. Almost constant number of Zn atoms are obtained from CB reconstruction, however, for SB reconstruction for the first arrow near the edge shows systematic increase in atoms number.

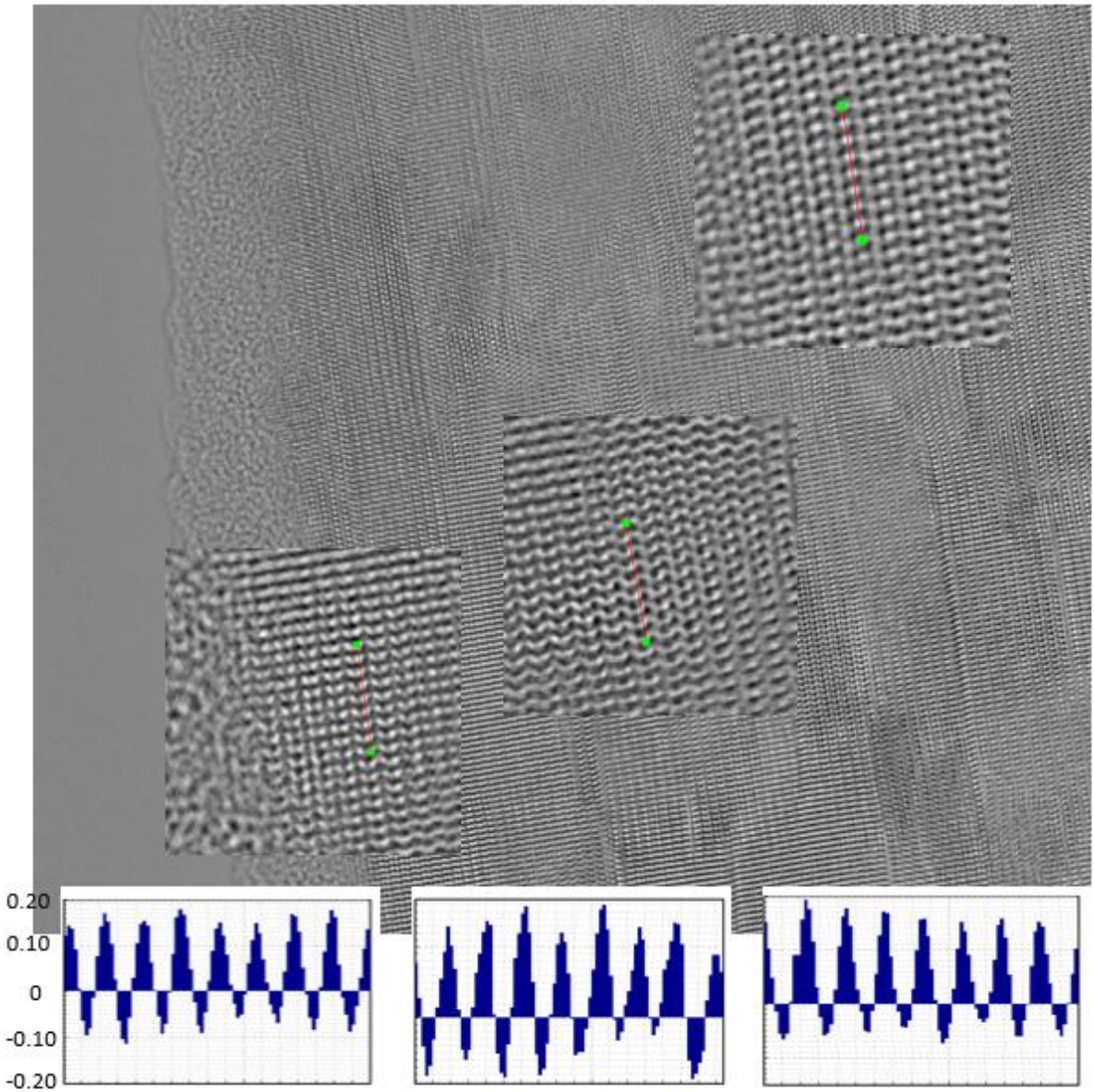

**Figure 9.** Reconstructed phase image of HRTEM image series for different thickness regions. Throughout the sample area almost constant phase and atom number 1 are obtained.

# Supplementary Document

# Quantitative atom counting of Zn and O atoms by atomic resolution off-axis and in-line holography

U. Bhat[1], and R. Datta[1]

[1]*International Centre for Materials Science, Chemistry and Physics of Materials Unit, Jawaharlal Nehru Centre for Advanced Scientific Research, Bangalore 560064, India.*

## S1. HRTEM image before and after digital aberration correction

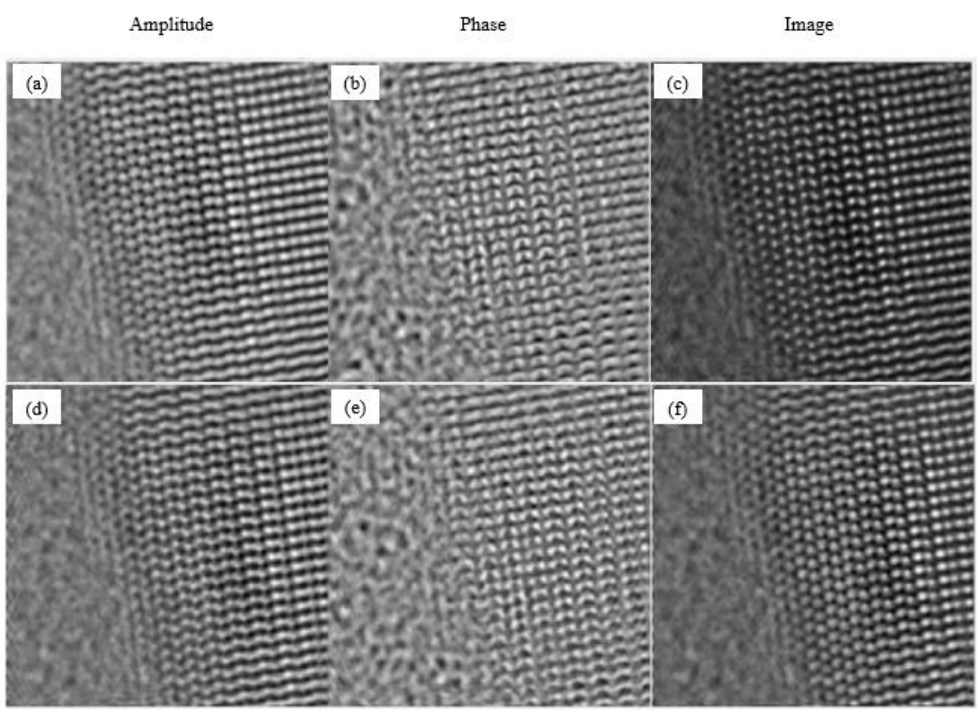

Fig S1. Example of HRTEM reconstructed amplitude, phase and image (a) - (c) before and (d)-(f) after digital aberration correction. Cs=-35µm, Δf=2nm are used for the aberration correction.

## S2. Phase contrast transfer function (PCTF):

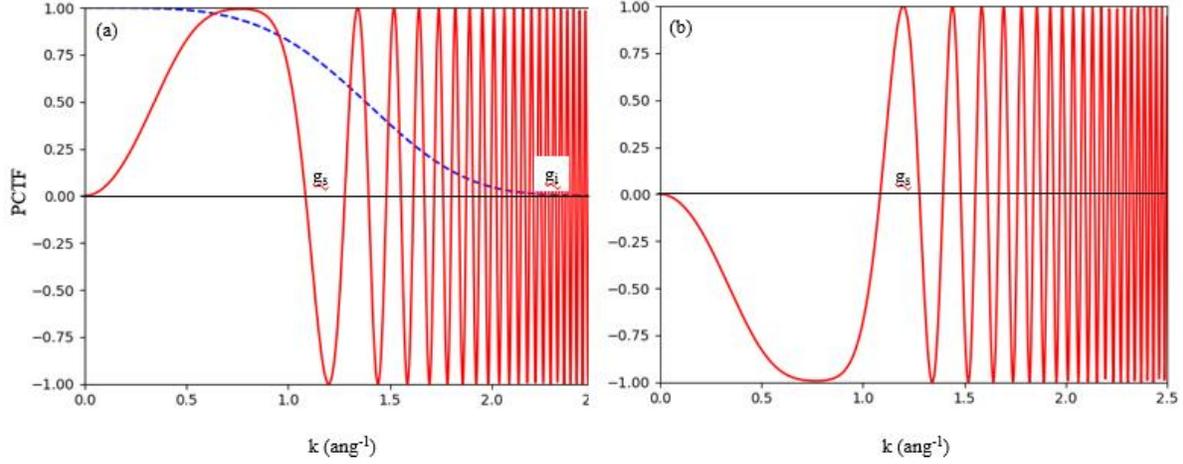

Figure S2.(a) PCTF function at 300 kV under optimum lens parameters, Cs=-35µm Δf=8nm with the envelope function (blue dotted line) corresponding to spread in defocus 1nm. The point resolution and information limits are marked as $g_s$ and $g_i$ respectively. The positive PCTF gives negative phase contrast or white atom contrast. (b) PCTF function corresponding to positive phase contrast or dark atom contrastwith Cs=35um, Δf=-8nm.

S 3. **On the phase detection limit**

In off-axis holography, the phase detection limit has been discussed by Lichte, [Ref. 36] and is given by

$$\sigma_\varphi = \frac{\sqrt{2e}}{p\sqrt{V^2 j_0 \tau STE(u_c)}} \qquad (S1)$$

The above equation can be written as

$$\sigma_\varphi = \frac{\sqrt{2}}{pV\sqrt{N/p^2 STE(u_c)}} \qquad (S2)$$

Where $N = \frac{1}{e}j_0 p^2 \tau$

N is the number of electrons/nm$^2$, $j_0$ is the mean current density in the detector plane, $p^2$ is the area of reconstruction, and $\tau$ is the exposure time. $STE(u_c)$ is the signal transfer function of the CCD camera.

The three important parameters in the above equations are, N, p and V. The fundamental limit in phase detection is governed by the shot noise or stochastic impacts of single electrons, due to probabilistic nature of the electron wave. This is given by

$$\sigma_\varphi = \sqrt{\frac{2c}{V^2 N}} \qquad (S3)$$

The fundamental phase detection limit improves with the increasing electron dose N. Lichte shown that for V=0.4, STE=0.8, and N=9000/nm$^2$ $\sigma_\varphi$ is 0.0314rad for $p^2$=1nm$^2$. However, there is almost no changes in the phase detection limit by improving contrast up to 0.8. rather decreases with further increase in contrast.

In our experimental holograms, acquired in Berlin, the average electron dose is 16*10$^6$ /nm$^2$, V=15%. Therefore, for a reconstructed area $p^2$=100nm$^2$(512 X 512), $\sigma_\varphi = 0.00023$ rad.

Lehmann has modified the equation S3 to incorporate the effect of Cs and smallest area of reconstruction (w$_{hol} \geq$ 4psf). Lehmann has shown improvement in the phase detection limit by factor of 4 using Cs corrected microscope. The minimum area of reconstruction for the present data is approximately $(4 \times 0.8)^2$ =(0.32nm)$^2$, where 0.8 Å is the point resolution of the microscope. Thus, the phase detection limit for the smallest area of reconstruction, correspondind to the present data is 0.007365rad. Theoretical model suggest, the change in peak (mean) phase due to incremental change in Zn and O atoms in the atomic column are 0.284 (0.122)and 0.1098 (0.052) rad respectively. The mean phase has been calculated for an inner and outer cut off potential 10-50pm for Zn atom. Therefore, it is possible to count the incremental Zn and O atom in the atomic columns of the ZnO from the present atomic resolution holography data irrespective of area of reconstruction. In this context, Cooper and Voelkl improved the phase detection limit to 0.001 and $2\pi/1000$ (0.00628) by long exposure and multiplicity of holograms along with bi-prism and sample drift correction, respectively [37,38]. However, none of the latter two cases above used double bi-prism set ups which eliminates Fresnel fringe and improves the phase detection limit significantly. However, there is another limit posed by reconstruction methods where standard deviation in vacuum phase value poses experimental phase detection limit [see section III. of the manuscript].

The calculated phase detection limit as a function of electron dose N, contrast V and p are shown in figure S3. The dashed vertical lines are marked corresponding to the current experimental parameters.

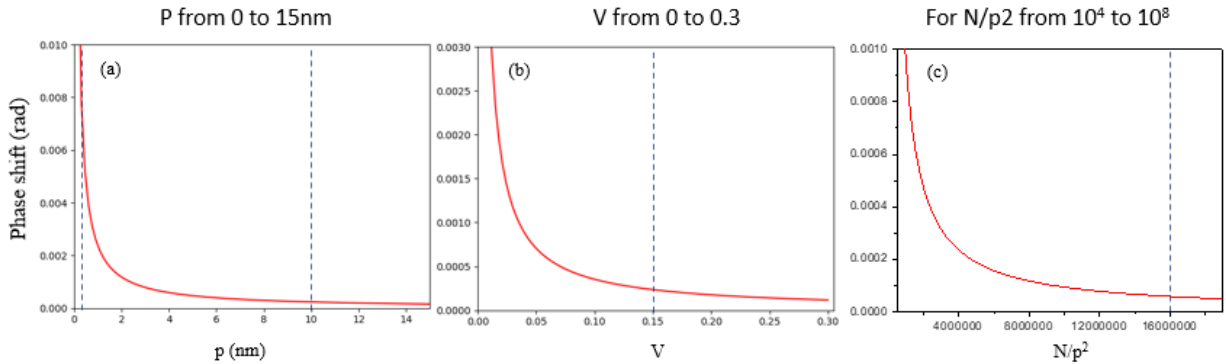

Figure S3. dependence of theoretical phase detection limit on area of reconstruction ($p^2$), visibility and electron dose are given.

S 4. **Experimental and theoretical phase shift value of B and N by in-line holography**

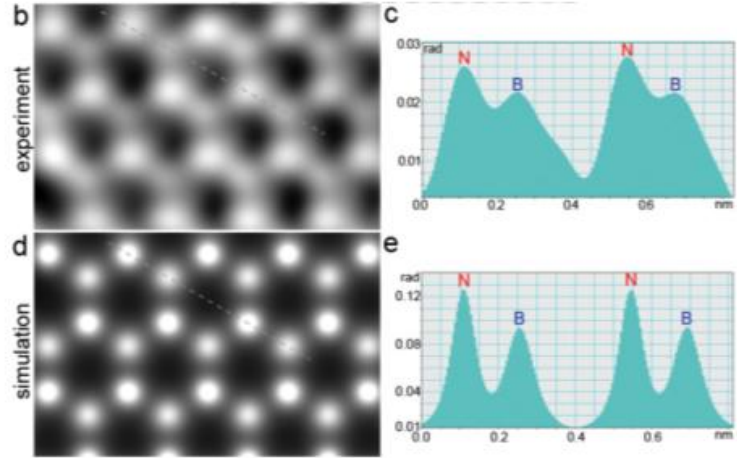

Fig S4. For a BN monolayer, the peak phase shift value of B and N are 0.022/0.09 and 0.026/0.13 with the difference 0.004/0.04 rad by experiment and simulation. Difference between the simulation and experimental values are because of Stobb's factor [39].

S 5. **Atomic potential model**

The charge distribution and corresponding atomic potential is calculated by Hartree-Fock procedure [Ref. 19]. The atomic potential in 3D is given by

$$V_a(x,y,z) = 2\pi^2 a_0 e \sum_{i=1}^{3} \frac{a_i}{r} \exp(-2\pi r \sqrt{b_i}) + 2\pi^{5/2} a_0 e \sum_{i=1}^{3} c_i d_i^{-3/2} \exp(-\pi^2 r^2 / d_i)$$

(S4)

with $r^2 = x^2 + y^2 + z^2$

Where, $a_0$ is the Bohr radius, $a_i, b_i, c_i, d_i$ are the parameterized coefficients.

Then the mean phase shift in the absence of dynamical scattering is calculated by the equation,

$$\Phi(x,y) = \sigma \int V(x,y,z)dz \qquad (S5)$$

Where $\sigma = \frac{\pi}{\lambda E}$ is the interaction parameter with wavelength $\lambda$ and accelerating voltage $E$ [44].

The projected atomic potential integrated along z-direction can be calculated from the equation above and is given in the main text equation (3).

The atomic potential is asymptotic and has a singularity at the center of the atom. Various resolution limiting factors such as diffraction limit, thermal vibration, aberration of the microscope result in measurable peak phase value in the phase image of the atom. The phase image of the atom can be approximated to a Gaussian function. Gaussian function is parameterized by the peak height and the full width half maxima (FWHM). Therefore, the reference phase shift can be considered either based on the peak value of the phase or the mean value of the phase. The mean value will depend on both the peak value and FWHM of the phase distribution function. The mean value of the phase can be calculated by integrating three-dimensional atomic potential between the two limits and dividing with the volume. The mean phase values calculated for different inner and outer cut-off values are given in table S1.

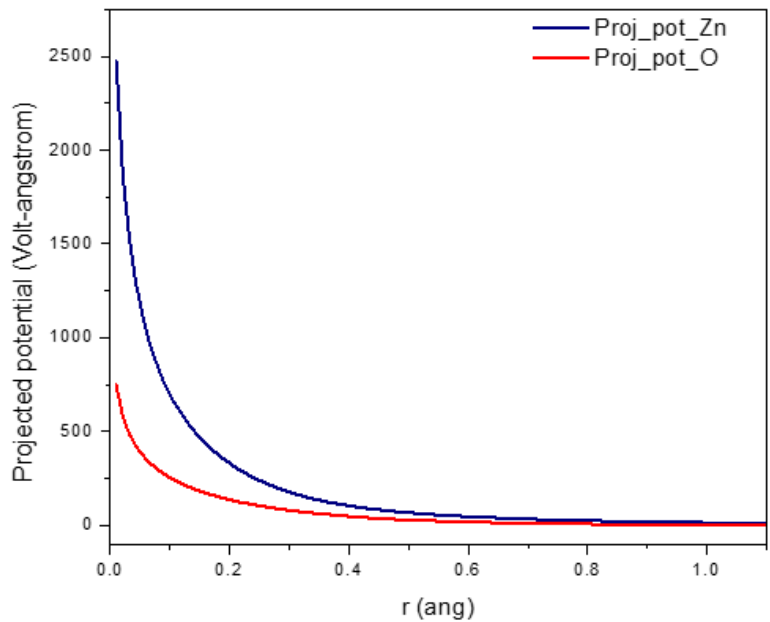

Figure S5. Projected potential of Zn and O atoms calculated using equation (3)

| Inner-Outer cut off (pm) | Mean phase shift (rad) | |
|---|---|---|
| | Zn | O |
| 0.1-100 | 0.056 | 0.022 |
| 1-100 | 0.054 | 0.022 |
| 5-100 | 0.054 | 0.019 |
| 10-100 | 0.051 | 0.021 |
| 5-50 | 0.133 | 0.055 |
| 5-60 | 0.107 | 0.045 |
| 5-80 | 0.073 | 0.031 |
| 10-50 | 0.122 | 0.052 |
| 10-60 | 0.099 | 0.043 |
| 10-80 | 0.069 | 0.030 |

Table S1. Table of mean phase shift with same outer cut off & varying inner cut off, same inner cut off with varying outer cut off for Zn and O atoms

S 6. **Scattering factor**

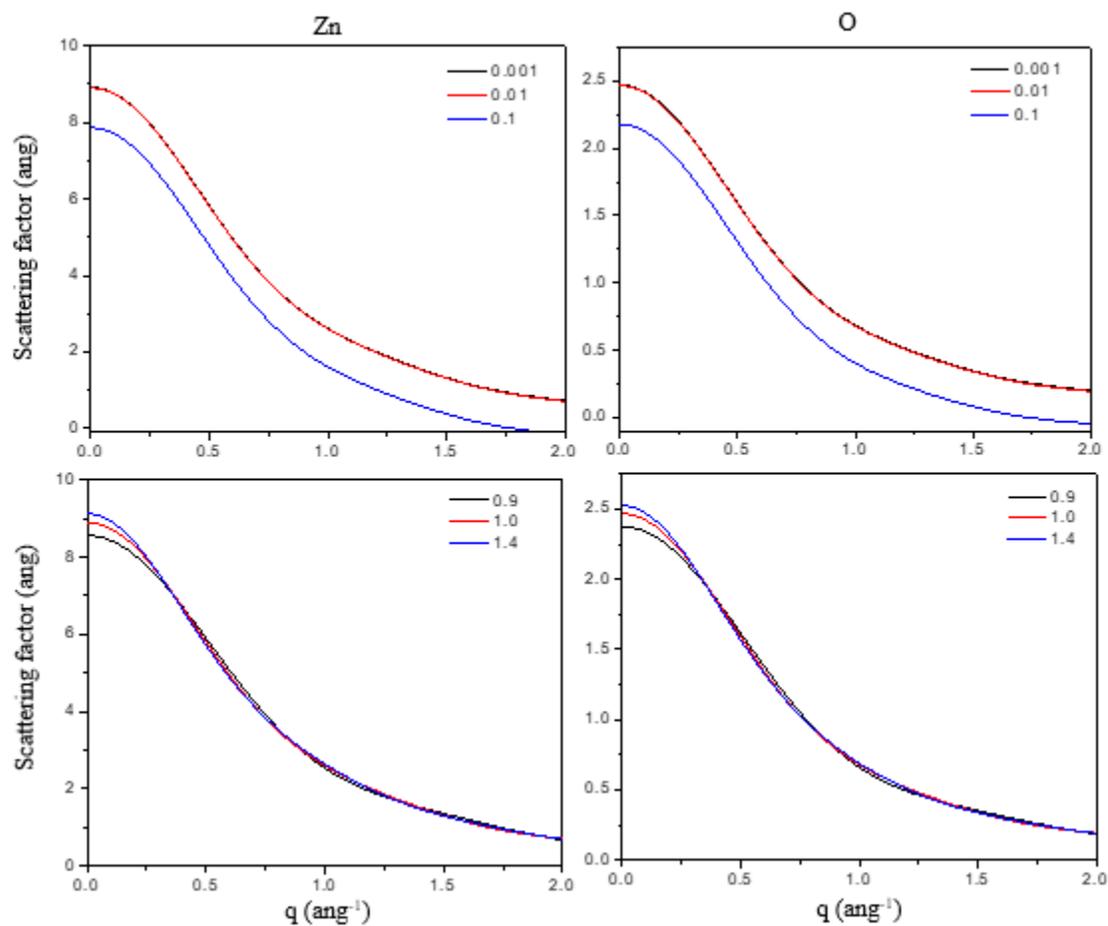

Figure S6. The variation of feq (according to Moliere) with inner and outer bound of the potential. There is no change in feq by changing inner cut off from 0.001 to 0.01 Å(for the same outer cut of 1 Å) but changes to inner cut off of 0.1ang for both Zn and O. On the other hand, keeping the inner cut off fixed(0.01 Å), there is only change in amplitude at small scattering angle (<0.25 Å$^{-1}$)by changing the outer cut-off.

S 7. **Peak phase as a function of cut-off**

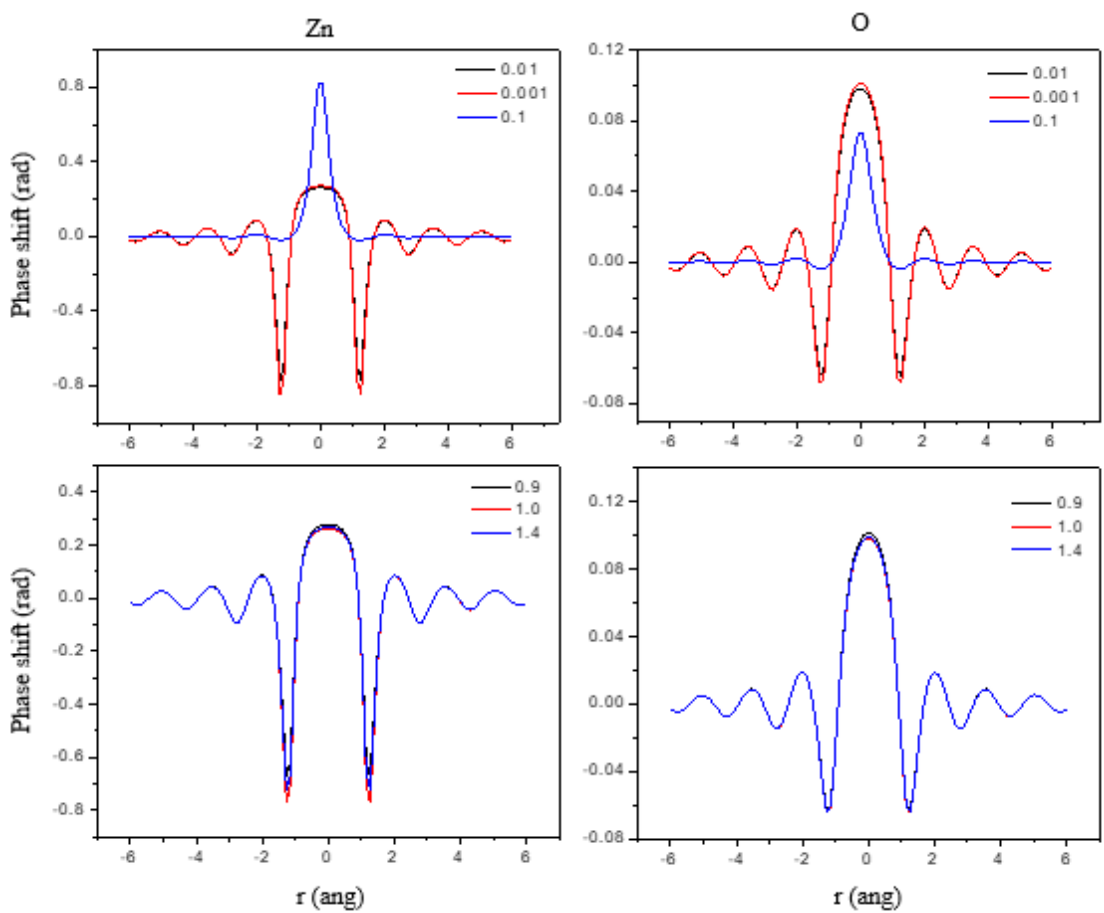

Figure S7. (a) & (c) The peak phase of Zn and O atom as a function of inner cut-off (fixed outer cut-off 1 Å) and (b) & (d) as a function of outer cut-off ( fixed inner cut-off, 0.01Å). The peak phase value does not change with the outer cut off potential from 0.9 to 1.4 Å. Peak phase values also do not change for inner cut-off of 0.001 and 0.01 Å but changes significantly for 0.1 Å.

S 8. **Peak intensity as a function of cut-off**

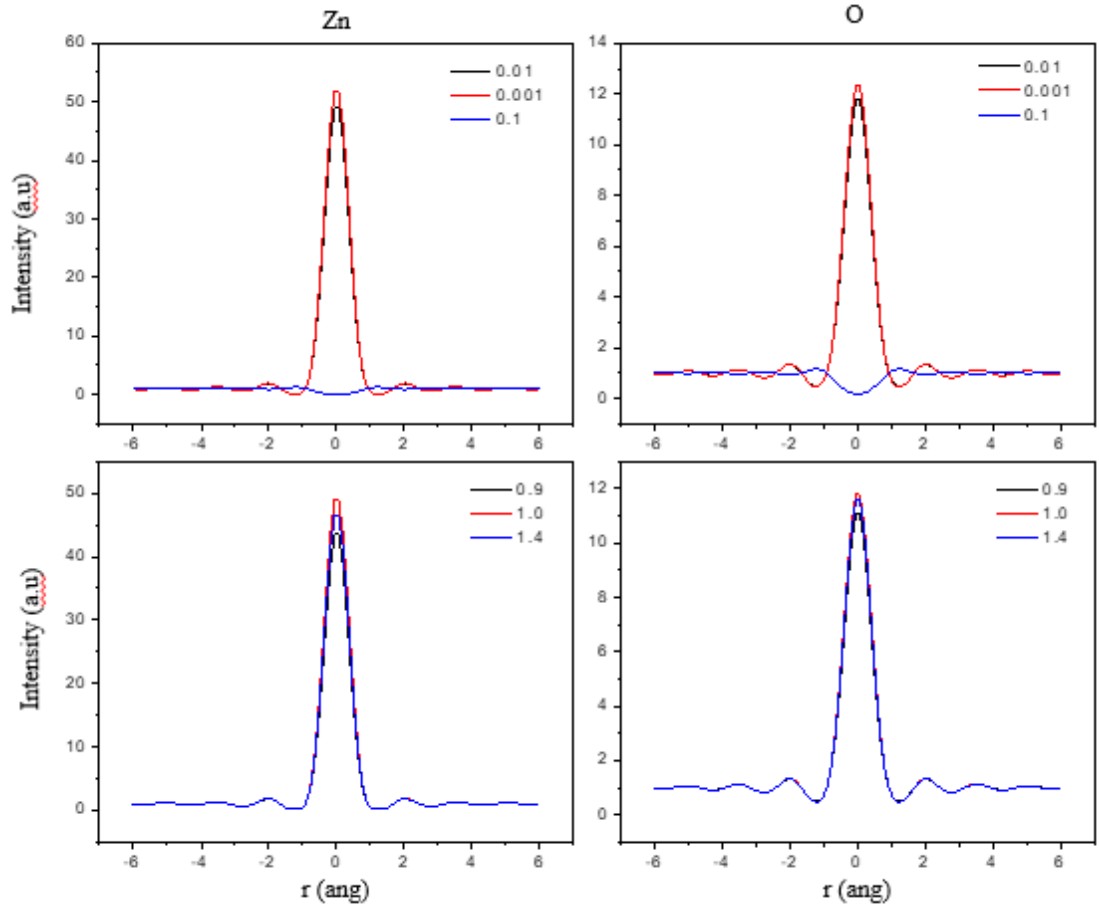

Figure S8. Corresponding intensity plots of Zn and O atoms, similar trend is observed in the intensity plot also with variation of inner and outer cut off potential.

S 9. **Phase shift and Intensity for different k**

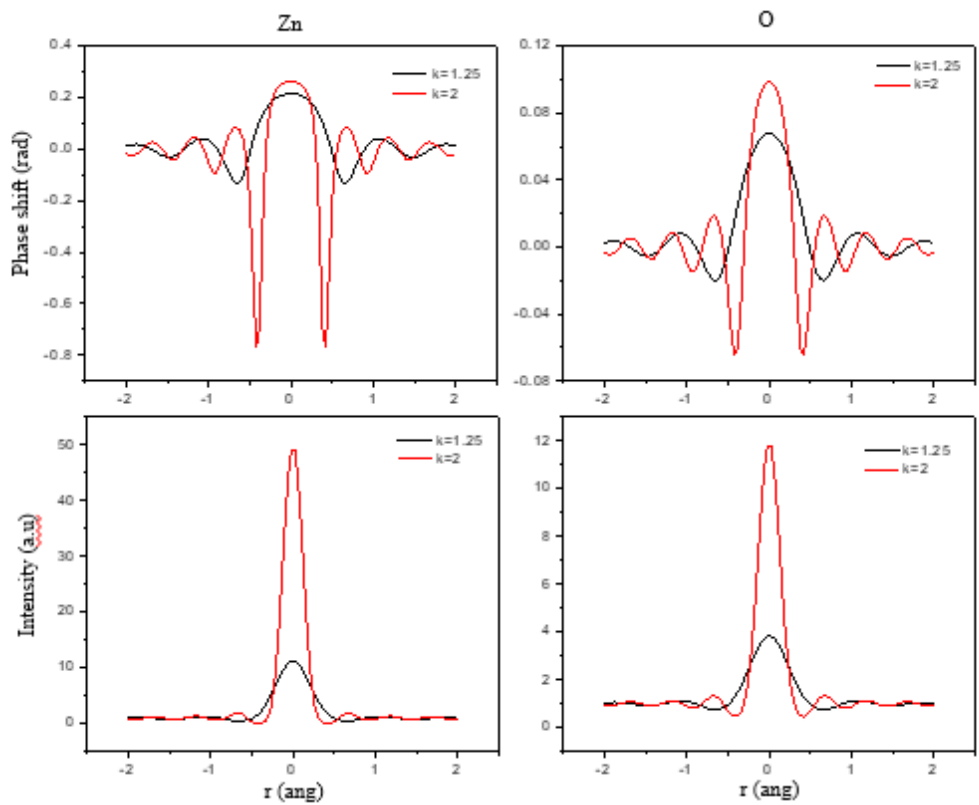

Figure S9. Phase shift and intensity plot of Zn and O atoms for k=1.25 and 2 Å$^{-1}$. The inner and outer cut off are 0.01 and 1 Å respectively.

S 10. **Reconstructed wave function**

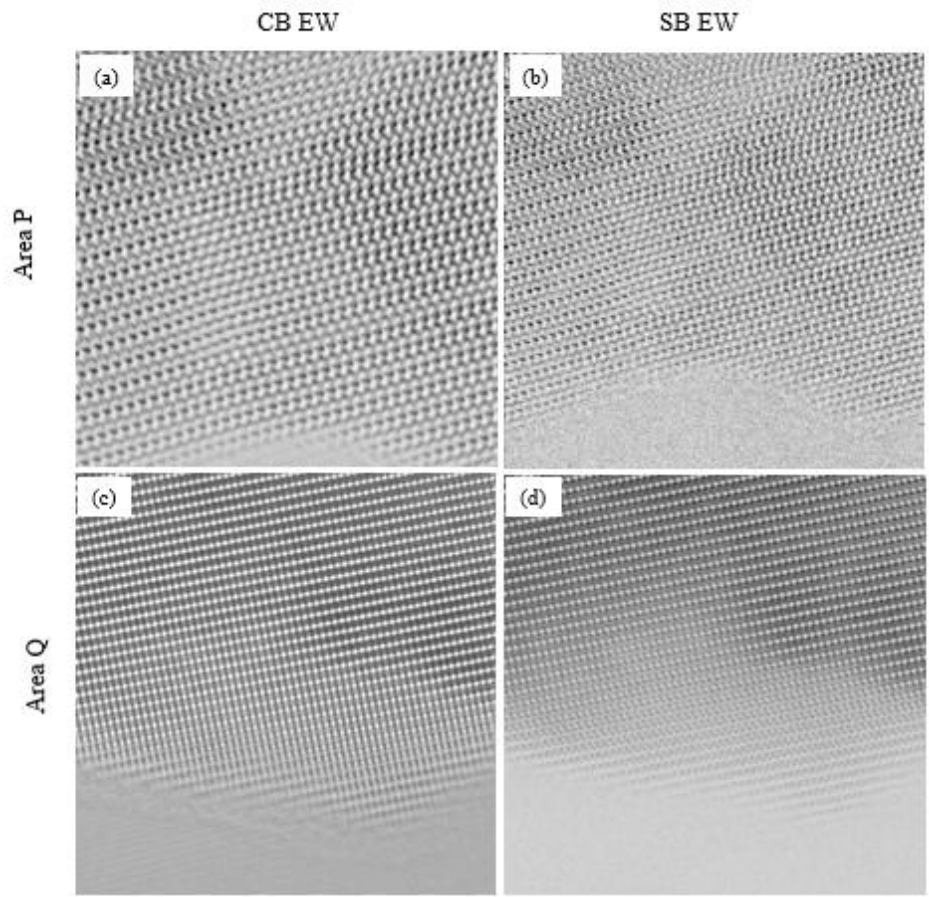

Figure S10. Reconstructed wave function of area P (a) and (b) and area Q (c) and (d) both from central and side band.

**Reference for supplementary**

[44] Williams D.B., Carter C.B. High-Resolution TEM. In: Transmission Electron Microscopy. Springer, Boston, MA. (2009) 486. **ISBN** 978-0-387-76501-3.